\documentclass{ws-procs975x65}

\begin{document}

\newcommand\be{\begin{equation}}
\newcommand\ee{\end{equation}}
\newcommand\bea{\begin{eqnarray}}
\newcommand\eea{\end{eqnarray}}
\newcommand\bseq{\begin{subequations}} 
\newcommand\eseq{\end{subequations}}
\newcommand\bcas{\begin{cases}}
\newcommand\ecas{\end{cases}}
\newcommand{\p}{\partial}
\newcommand{\f}{\frac}

\title{Generic Evolutionary Quantum Universe}

\author{Marco Valerio Battisti}

\address{ICRA - International Center for Relativistic Astrophysics \\
Physics Department (G9), University of Rome ``La Sapienza''\\ P.le A. Moro 5, 00185 Rome, Italy \\
E-mail: battisti@icra.it}

\author{Giovanni Montani}

\address{ICRA - International Center for Relativistic Astrophysics \\
Physics Department (G9), University of Rome ``La Sapienza''\\ P.le A. Moro 5, 00185 Rome, Italy \\
ENEA C.R. Frascati (U.T.S. Fusione), Via Enrico Fermi 45, 00044 Frascati, Rome, Italy \\
E-mail: montani@icra.it}

\begin{abstract}
We consider a Schr\"odinger quantum dynamics for the gravitational field
associated to a generic cosmological model and then we solve
the corresponding eigenvalue problem. We show that, from a phenomenological point of view, 
an Evolutionary Quantum Cosmology overlaps the Wheeler-DeWitt approach.
\end{abstract}

\bodymatter

\section{Introduction}\label{aba:sec1}
The canonical approach in Quantum Gravity is characterized by the so-called {\it frozen formalism}, i.e. the absence of a time evolution for the wave functional\cite{DeW67,Ku81}. It has been proposed\cite{Mo02,MeMo04a} that such feature disappears as soon as the impossibility of a physical slicing without frame fixing is recognized for a quantum spacetime. In this work\cite{BaMo06} we start with a Schr\"odinger dynamics for the gravitational field using Planck mass particle as a ``clock'' for the system; and we will analyze the meaning of the corresponding spectrum (we deal with a new energy density) in the framework of a generic inhomogeneous Universe.

\section{Evolutionary Quantum Gravity}
In this section we briefly analyze the implication of a Schr\"{o}dinger formulation of the quantum dynamics for the gravitational field.
We require the theory to evolve along the spacetime slicing so that $\Psi=\Psi(t,\left\{h_{ij}\right\})$; so the quantum evolution is governed by a smeared Schr\"{o}dinger equation
\be
i\p_t \Psi=\hat{\mathcal{H}}\Psi\equiv\int_{\Sigma_t^3}d^3x\left(N\hat{H}\right)\Psi
\ee
being $\hat{H}$ the super-Hamiltonian operator and $N$ the lapse function. If we now take the right expansion for the wave functional, the Schr\"{o}dinger dynamics is reduced to an eigenvalues problem of the form
\be\label{eipro}%
\hat{H}\chi=\epsilon\chi,\qquad \hat{H_i}\chi=0,
\ee
which outlines the appearance of a non zero super-Hamiltonian eigenvalue. 

Is not difficult to show that the classical limit (in the sense of WKB approximation) of the above model is characterized by the appearance of a new matter contribution, which admits the following energy density:  
\be\label{enden}%
\rho\equiv T_{00}=-\f{\epsilon(x^i)}{2\sqrt h}, \qquad h=\det h_{ij}.
\ee     
The explicit form of (\ref{enden}) is that of a dust fluid co-moving with the slicing 3-hypersurfaces, i.e. we deal with $T_{\mu\nu}=\rho n_\mu n_\nu$.

\section{The Generic Quantum Universe and its Spectrum}
We now apply the Schr\"{o}dinger approach to a Quantum Universe that has to be described by a generic inhomogeneous model\cite{BKL82}, which has a dynamics summarized, asymptotically to the Big-Bang, by the following variational principle\cite{BeMo04}
\be
\delta S=\delta\int_{\Sigma_t^3\times\Re} dt d^3x(p_a \p_t q^a -NH)
\ee
where adopting Misner-like variables $R$, $\beta_{\pm}$\cite{Mis69a} ($R$ is the scale factor and $\beta_{\pm}$ describes the anisotropies), the super-Hamiltonian has the structure
\be\label{eigenpro}%
H(x^i)=\kappa \left[-\f{p_R^2}R+\f{1}{R^3}\left(p^2_+ +p^2_-\right)\right]+\f{3}{8\pi}\f{p^2_\phi}{R^3}-\f{R^3} {4\kappa l^2_{in}} V(\beta_\pm)+R^3(\rho_{ur}+\rho_{pg}),
\ee
where $\kappa=8\pi l_P^2$. We have added to the dynamics of the system an ultrarelativistic energy density ($\rho_{ur}=\mu^2/R^4$), a perfect gas contribution ($\rho_{pg}=\sigma^2/R^5$) and a scalar field $\phi$ (a free inflaton field).  

Performing the canonical quantization of this model we obtain the following eigenvalue problem (\ref{eipro}), with the right normal ordering:
\be\label{H}%
\left\{\kappa \left[\p_R\f 1 R\p_R-\f 1 {R^3}\left(\p_+^2+\p_-^2\right)\right]-\f{3}{8\pi R^3}\p_\phi^2 -\f{R^3} {4\kappa l^2_{in}}V(\beta_\pm)+R^3(\rho_{ur}+\rho_{pg})\right\}\chi= \epsilon\chi.
\ee 
The appropriate boundary condition for this problem are: i) $\chi(R=0,\beta_{\pm},\phi)<\infty$ that relies on the idea that the quantum Universe is singularity-free and ii) $\chi(R\rightarrow\infty,\beta_{\pm},\phi)=0$ that ensures a physical behavior at ``large'' scale factor.

In order to study the previous eigenvalue problem we expand the wave function as $\chi(R,\beta_{\pm},\phi)=\int \theta_K(R)F_K(R,\beta_{\pm},\phi)dK$, and then performing an adiabatic approximation ($|\p_RF|\ll|\p_R\theta|$) we obtain the following
reduced problems:
\be\label{eqR}%
\kappa\f d {dR}\left(\f 1 R \f{d\theta}{dR} \right) + \left(\kappa \f{K^2}{R^3}+R^3(\rho_{ur}+\rho_{pg})-\epsilon\right)\theta=0,
\ee
\be\label{eqbeta}%
-(\p_+^2+\p_-^2+\f3{8\pi\kappa}\p_\phi^2)F+\f{R^6}{4\kappa^2 l^2_{in}}V(\beta_{\pm})F=K^2(R)F.
\ee  
The function $F$ is a plane wave as soon as we neglect the potential term in (\ref{eqbeta}) for same $R^\ast\ll 1$. The solution of (\ref{eqR}) is a series in $R$ multiplied by a Gaussian function peaked around $R=\epsilon l_P^2/16\pi$. Since we required the wave function to decay at large scale factor $R$ we have to terminate the series and obtain the spectrum of the super-Hamiltonian:
\be\label{spee}%
\epsilon_{n,\gamma}=\f{\sigma^2}{l_P^2(n+\gamma-1/2)},
\ee
so that the ground state $n=0$ eigenvalue, for $\gamma<1/2$, is negative; therefore is associated via (\ref{enden}) to a positive dust energy density.

\section{Phenomenology of the Dust Fluid}
In order to analyze the cosmological implication of this new matter contribution, we have to impose a cut-off length in our model, requiring that the Planck length $l_P$ is the minimal physical length accessible by an observer ($l\geq l_P$). So, from the thermodynamical relation for the perfect gas, we obtain a constraint on the $\rho_{pg}$ and then on the super-Hamiltonian eigenvalue:
\be
l^3 \equiv \f V {\cal{N}}=\f{3}{2}\f{l_P}{\rho_{pg}\lambda^2}\geq l_P^3 \qquad \Rightarrow \qquad \rho_{pg}\leq\mathcal{O}(1/l_P^4),
\ee      
where $l$ is the length per particle and $\lambda$ the corresponding thermal length ($\lambda=l_P$). Therefore we get $\sigma^2\leq\mathcal{O}(l_P)$ and so $|\epsilon_0|\leq\mathcal(1/l_P)$: {\it the spectrum is limited by below}.

The contribution of our dust fluid to the actual critical parameter is
\be
\Omega_{dust}\sim \f{\rho_{dust}}{\rho_{Today}}\sim\mathcal{O}\left( 10^{-60}\right).
\ee
Such a parameter is much less then unity and so no phenomenology can came out (today) from our dust fluid. In other words an Evolutionary Quantum Cosmology overlaps the Wheeler-DeWitt approach. Finally we face the question of the classical limit of the spectrum in the sense of large occupation numbers $n\rightarrow \infty$. As we can see from (\ref{spee}) the eigenvalue approaches zero as $1/n$. 
Therefore for very large $n$, our quantum dynamics would overlap the Wheeler-DeWitt approach.

\end{document}